\documentstyle[12pt]{aipproc}
\begin{document}
\title{Experiments on Sonoluminescence: Possible Nuclear and QED Aspects
and  Optical Applications}

\author{V.B.\,Belyaev$^{(1)}$, Yu.Z.\,Ionikh$^{(2)}$, M.B.\,Miller$^{(3)}$,
A.K.\,Motovilov$^{(1)}$, A.V.\,Sermyagin$^{(3)}$, A.A.\,Smolnikov$^{(1,4)}$,
Yu.A.\,Tolmachev$^{(2)}$}%

\address{\small $^{(1)}$Joint Institute for Nuclear Research, Bogolubov Laboratory
of Theoretical Physics, \mbox{141980 Dubna, Moscow reg., Russia}\\
$^{(2)}$Institute of Physics of the St.\,Petersburg State University,
 \mbox{198904 Peterhof, St.\,Petersburg reg., Russia}\\
$^{(3)}$Institute of Physical and Engineering
Problems\thanks{E-mail: iftp@dubna.ru}
 \mbox{P.\,O.\,Box\,39, 141980 Dubna, Moscow reg., Russia}\\
$^{(4)}$Institute for Nuclear Research, Baksan Neutrino Observatory,
  \mbox{117312 Moscow, Russia}}

\maketitle

\begin{abstract}
Experiments aimed at testing some hypothesis about the nature
of Single Bubble Sonoluminescence are discussed.
A possibility to search for micro-traces of thermonuclear neutrons
is analyzed, with the aid of original low-background neutron
counter operating under conditions of the deep shielding from
Cosmic and other sources of background. Besides, some signatures
of QED-contribution to the light emission in SBSL are under
the consideration, as well as new approaches to probe a
temperature inside the bubble. An applied-physics portion of
the program is presented also, in which an attention is being
paid to single- and a few-pulse light sources on the basis of SBSL.
\end{abstract}

\section*{Introduction}

A lack of a complete explanation of some
unusual characteristics of sonoluminescence came to be a source of
a few exotic suggestions about its nature. For our
studies we selected the most intriguing ones that are somehow
in line with a general
scientific stream  in
Dubna research center.
Search for predicted nuclear and quantum electrodynamic effects in
sonoluminescence is an aim of described experiments that are  now in
a stage of preparation.

\section*{Experimental approaches}

\paragraph*{Hot-plasma hypotesis.} It was predicted that very high
temperatures are possible at the
moment of a collapse of the sonoluminescencing bubble. Under
some special mode of upscaling sonoluminescence (a strong short
pressure pulse should be added to the ultrasound standing wave)
the temperature presumably
reach a level of observable traces of
thermonuclear fusion in the D+D system.
 The additional
short pressure
pulse with a magnitude of several bars are to be synchronous
with the sonoluminescence
flash. If the system contains deuterium dissolved in heavy
water (D$_2$O)  a neutron yield is expected. A value about 0.1
nph is predicted for some optimal conditions~\cite{Moss}. Measurements of
this low neutron rate are planned to be
done by means of a triple-coincidence method using an
original
neutron counter. The neutron spectrometer was designed taking into account
requirements for
minimizing the $\gamma$-ray and random coincidence backgrounds~\cite{12}. It is a
calorimeter based on a liquid organic scintillator-thermalizer with
$^3$He
proportional counters of thermalized neutrons distributed uniformly over the
volume. The energy of thermalized neutrons is transformed into light signals
in a scintillation detector. The signals from proportional counters provide
a `neutron label' of an event.
The triple
coincidences are to be sorted by a following algorithm:

{\it(signal from the sonoluminescence-light flash) \&
(the scintillator flash in moderator) \&
(the signal from the $^3$He counter)}.

The measurements are supposed to be performed
in the underground laboratory of the Baksan Neutrino Observatory of
the Institute for Nuclear Research of  Russian  Academy of
Sciences, Caucasus. A shield from cosmic rays in this Lab is
about 5,000\,m of w.\,e. %
Under these
conditions a sensitivity of
about 0.01 nph can be reached for about a three-month measurement cycle. The main
components of necessary devices and equipment are already in our
disposal, including the sonoluminescence devices and the neutron
counter.

The system of intensive pressure pulsing is under construction.
Certainly, many efforts
are necessary to modify the experimental setup and accommodate
it for the measurements at Baksan. To reach the most possible
sensitivity in these experiments it would be reasonable to use a
so-called few-bubble-sonoluminescence (FBSL) regime  when several single
bubbles are trapped within a higher harmonic modes of acoustic
resonator  as it was reported in~\cite{Geisler}.
We have observed
concurrently lighting SL-bubbles in the second harmonic
under some special
boundary conditions which have yet to be specified~\cite{Belyaev}.

\paragraph*{QED-hypotesis.} Another idea connects the sonoluminescence
to the energy of zero vibrations of the vacuum (Casimir
energy)~\cite{13}.  To test this idea the following two types of
experiments are designed.

{\it(1) Transforming a short wave part of the sonoluminescence
spectrum to the region of $\lambda$ higher than the water
absorption edge.}  To this end, certain specific
luminofores should be selected, and among them perhaps tiny dispersive
powder of crystal r\"ontgenoluminofores would be promising as an
interaction with lattice is essential in this case. Certainly,
possible influence of those suspensions on cavitation properties
of water
have to be clarified beforehand.
If the hyposesis on the
vacuum-fluctuation nature of the sonoluminescence phenomenon is
valid then no short wave emission with $\lambda<200$\,nm is
expected.

{\it(2) Angular-correlation measurements of the coincident photons in the
vicinity of 180$^\circ$ .} The QED model for the
sonoluminescence predicts the emission of time-correlated pairs
of photons flying away in opposite directions.
\vskip 0.5cm
Some other experiments are considered also. In particular,

{\it (1) Direct testing the so-called dissociation
hypothesis (DH)  of the
sonoluminescence.} According to DH, when the
stable single bubble sonoluminescence conditions have been created,
the inert gas alone remains inside the bubble. Due to high temperature
inside the bubble all other components (nitrogen and
oxygen) undergo the intensive chemical interaction with each
other and with water. This results in nitrogen oxides
(NO, NO$_2$) and NH$_3$~\cite{14}.
At present only indirect evidence for the DH has been
found~\cite{15,16}. For direct measurement of the
 products of dissociation  a small SL cell,
 completely closed to the atmosphere, and containing a relatively
 small volume of water, will be needed. Long-time runs can be
 accomplished via computer-controlled monitoring of the system ~\cite{MATULA PC}.

   {\it (2) Measurements of spatial distribution of the light in water.}
Measurements of the time-averaged spectral distribution of the
radiation emitted by the bubble will be done
in the presence of the
luminofore
additives that will be used in the procedure of
spectrum-trasformation experiments. The aim of the experiments is to infer the
source brightness
in the short UV range by means of  taking
into account the diffusion of UV emission in the water solutions or suspensions of the
above luminofores, and, by
comparing them with
predictions of  quasi-stationary thermal source model,
estimates of   plasma temperature are expected to be obtained~\cite{17}.

   {\it (3) Study of near-IR spectrum of the emission of Xe-doped bubble.}
 We will try to search for spectral lines of Xe similar to distinctive line
 emissions
  observed in high pressure xenon lamps.

\section*{Development of new type of super fast pulse light sources}

One of the most remarkable features of SBSL flashes is
their brief duration.
The most recent measurements show that
many parameters,
such as the nature and concentration of gases,
temperature, pressure, resonance performances of the SL instrument,
etc., influence the temporal and other properties. For example,
larger flasks operating at lower frequencies, cause the bubble to
emit more light~\cite{18}. It is important that
the light pulse duration remains the same within the limit of a few
picosecond for
whole  wavelength range .
 \vskip 0.5cm

Goals of this part of experiments are:

(1) Studying parameters controlling the duration of light
pulses and other temporal parameters of SBSL radiation.
(2) Investigation of correlation in intensity of flashes.
In this experiment the statistical properties of the intensity
of the light flashes will be studied to determine the short-term
 and long-term aspects of SL. The synchronicity between
flashes has
been shown to be remarkably high~\cite{19}. It is interesting whether the
intensity distribution is as narrow, as well.
  (3) Development of SL-light sources of single light pulses.
 To this end the Kerr cells will be applied. Regular and relatively
 rare repetition of the flashes makes possible using the
 photo-shutters with a limited temporal resolution to obtain
 the single pulses with temporal performances determined by
 primary SBSL-properties.
 (4) Development of methods  to generate various series
  of the single light
 pulses.
 (5) On this basis, developing simple, inexpensive light
sources for physical research, fist of all for the
time-resolution calibration of fast photodetectors (PMT and the
like).

\section*{Acknowledgments}
The authors are grateful to Prof.\,L.\,A.\,Crum and Dr.\,T.\,J.\,Matula
for very fruitful discussions, and to Prof.\,W.\,Lauterborn
for his interest to this program.\\
\par
This work was supported in part by the Russian Foundation for Basic
Research, Grant No.\ 98--02--16884.

\end{document}